\documentstyle[preprint,aps]{revtex}
\font\titlefont=cmbx10 scaled \magstep4

\begin{document}
\draft
\tighten
\begin{flushright}
\vspace*{-2cm}
quant-ph/0112056 \\
\vspace*{.2cm}
\end{flushright}

\begin{center}
{\titlefont FLUCTUATIONS OF THE RETARDED}\\
\vspace*{0.4cm}
{\titlefont VAN DER WAALS FORCE }\\
\vspace*{1cm}
Chun-Hsien Wu \footnote{e-mail: wu@cosmos.phy.tufts.edu} \\
Institute of Cosmology, Department of Physics and Astronomy\\
Tufts University\\
Medford, MA 02155, U. S. A.
\vspace*{0.2cm}
\\ Chung-I Kuo \footnote{e-mail: cikuo@mail.scu.edu.tw}\\
Department of Physics\\
Soochow University\\
Taipei, Taiwan, Republic of China\\
\vspace*{0.2cm}
and
\vspace*{0.2cm}
\\L. H. Ford \footnote{e-mail: ford@cosmos.phy.tufts.edu} \\
Institute of Cosmology, Department of Physics and Astronomy\\
Tufts University\\
Medford, MA 02155, U. S. A.
\end{center}
\vspace*{-1cm}

\begin{abstract}
The retarded Van der Waals force between a polarizable particle and a 
perfectly 
conducting plate is re-examined. The expression for this force given 
by
Casimir and Polder represents a mean force, but there are large 
fluctuations around this mean value on short time scales which are of 
the same order
of magnitude as the mean force itself. However, these fluctuations 
occur on  time 
scales which are typically of the order of the light travel time 
between the atom and the plate. 
As a consequence, they will not be observed in an experiment which 
measures the 
force averaged over a much longer time. In the large time limit, the 
magnitude of the mean squared 
velocity  of a test particle due to this fluctuating Van 
der Waals force
approaches a constant, and is similar to a Brownian motion of a test 
particle in
an thermal bath with an effective temperature. However the 
fluctuations are not 
isotropic in this case, and the shift in the mean square velocity 
components can even be negative. We interpret this negative shift to 
correspond to a reduction in the velocity spread of a wavepacket.
The force fluctuations discussed in this paper are special case of the more
general problem of stress tensor fluctuations. These are of interest in a 
variety of areas fo physics, including gravity theory. Thus the effects
of  Van der Waals force  fluctuations serve as a useful model for better
understanding quantum effects in gravity theory.
 
\end{abstract}
\vfill
\eject

\baselineskip=13pt 

\section{Introduction}
 
      The retarded Van der Waals forces between pairs of atoms and 
between
an atom and a perfectly conducting plate were first calculated by 
Casimir
and Polder\cite{CP}.  In the long distance limit, where the atoms may 
be 
described by a static polarizability, these forces may be interpreted 
as due 
to shifts in the vacuum energy of the quantized electromagnetic 
field. This 
is most clearly illustrated by the Casimir effect\cite{Casimir}, 
which may
be viewed either as the retarded Van der Waals force between a pair 
of 
perfectly conducting plates, or as the shift in the vacuum energy due 
to
the plates. It has recently been measured accurately 
\cite{Lamoreaux,Mohideen,Mohideen2}. Similarly, the 
Casimir-Polder 
force between a plate and an atom has been confirmed by a experiment 
by  Sukenik et al \cite{Sukenik}. Note that the large distance limit of the
theory can be applied to any polarizable particle, but not just an atom.

Because these forces have their origin in the vacuum fluctuations of 
the
electromagnetic field, it is perhaps not surprising that the forces 
themselves are fluctuating forces. The first discussion of the
force fluctuations was given by Barton\cite{Barton,Barton92},
who considered fluctuations of the Casimir force between plates. 
In this approach, one considers a spatial and/or time average of the 
force.
It is found that the fluctuations diverge in the limit that the 
averaging
time goes to zero. Further work along the same lines was done by 
Eberlein\cite{Eberlein}. Jaekel and Reynaud\cite{JR} have also 
discussed
Casimir force fluctuations, especially for accelerating mirrors, using
an approach based upon fluctuation-dissipation theorems. In this 
paper, 
we will consider the fluctuations of the force between
an atom and a perfectly conducting plate from an approach somewhat 
different
from that adopted by either of the above sets of authors. Our approach
is based upon the Langevin equation. The solution of this equation to
find the mean squared velocity of the particle involves a time integration
which introduces a natural averaging scale. We will show that this averaging 
is sufficient to yield finite results.

The problem addressed in the present paper can be viewed as a special case
of the larger problem of understanding the quantum fluctuations of the
stress tensor \cite{Kuo,WF99,WF00}. This problem is of interest for
a variety of reasons, ranging from radiation pressure noise in an
interferometer \cite{WF00}, to quantum fluctuations of spacetime
geometry driven by stress tensor fluctuations \cite{Kuo,WF01}. 

This paper is organized as follows: The Van der Waals force is 
reviewed in Sec.~\ref{sec: mean force} and then the force -force correlation 
function will be calculated in Sec.~\ref{sec: force correlation}. 
In Sec.~\ref{sec:average}, we use this correlation function to study
the velocity fluctuations of a test particle. Here it will be useful to
use a decomposition of the correlation function into three parts, 
 and to study the effect of each part individually.
Our results will be summarized and discussed in Sect.~\ref{sec:discuss}.

\section{The Van der Waals Force} \label{sec: mean force}

First, let us recall the result for the mean force. We assume that 
the atom can
be described as a point particle with a static polarizability 
$\alpha$.
Its interaction energy with a classical electromagnetic field, ${\bf 
E}$,
is
\begin{equation}
U=-{\alpha\over 2}\,{\bf E}^2.
\end{equation}
We will use Lorentz-Heaviside units with $\hbar = c =1$, but will 
restore
factors of $\hbar$ and $c$ in key results.
We now assume that the electromagnetic field is quantized, and that 
its
quantum state is such that $\langle{\bf E}\rangle =0$. However, 
$\langle{\bf E}^2\rangle \not=0$, and there is a mean force given by
\begin{equation}
\langle{\bf F}\rangle ={\alpha\over 2}\,{\bf \nabla}\,
                                            \langle{\bf E}^2\rangle.
                                                  \label{eq:force}
\end{equation}

     Quantities such as $\langle{\bf E}^2\rangle$ in the presence of
a plate may be calculated from the photon Hadamard function:
\begin{equation}
G_{\mu\nu'}\equiv G_{\mu\nu}(x,x') \equiv
{1\over 2}\,\langle A_\mu (x)\,A_\nu (x') + A_\nu (x')\,A_\mu 
(x)\rangle \, 
= G^{(0)}_{\mu\nu'} + {\tilde G}_{\mu\nu'}\, ,
\end{equation}
where
\begin{equation}
G^{(0)}_{\mu\nu'} = { {\eta_{\mu\nu}} \over {4\pi^2 D (x, x')} }
    \label{eq:G_0}
\end{equation}
is the Hadamard function for empty space, and
\begin{equation}
{\tilde G}_{\mu\nu'} = -{ {\eta_{\mu\nu} -2\,{\hat z}_\mu {\hat 
z}_{\nu'}} 
\over {4\pi^2 \tilde D (x, {\tilde x'})} }  \label{eq:Gtilde}
\end{equation}
is an ``image'' term due to the presence of the conducting 
boundary\cite{BM,F93}.
Here $\eta_{\mu\nu} = diag(-1,1,1,1)$ is the Minkowski metric, and
${\hat z}$ is the unit vector in the $z$ direction
\begin{equation}
{\hat z}^\mu=(0,0,0,1)\, .
\end{equation}
Furthermore, $D (x, x')$ is the squared geodesic distance between
$x$ and $x'$,
\begin{equation}
D (x, x') = -(t-t')^2 +(x-x')^2 +(y-y')^2 +(z-z')^2 \, ,
          \label{eq:D_0}
\end{equation}
and $\tilde D (x, {\tilde x'})$ is the corresponding distance between 
$x$
and the image point ${\tilde x'}= (t',x',y',-z')$,
\begin{equation}
\tilde D (x, {\tilde x'}) = -(t-t')^2 +(x-x')^2 +(y-y')^2 +(z+z')^2\,.
\end{equation}

     The vacuum expectation value of a product of
electric fields in the presence of a conducting plate is 
given by
\begin{equation}
\langle E_i\,E_{j'}\rangle =
\partial_0 \,\partial_{0'} \,G_{ij'}+
\partial_i \,\partial_{j'} \,G_{00'}-
\partial_0 \,\partial_{j'} \,G_{i0'}-
\partial_i \,\partial_{0'} \,G_{0j'}.    \label{eq:E2}
\end{equation}
We are using a notation in which unprimed indices refer to the 
spacetime
point $x$, and primed indices to the point $x'$. Thus, $\partial_0 =
\partial/\partial t$, $\partial_{j'} = \partial/\partial x^{j'}$, etc.
The quantity $\langle E_i\,E_{j'}\rangle$
is divergent in the coincidence limit, $x' \rightarrow x$,
but the divergent part does not contribute to the force in 
Eq.~(\ref{eq:force}). For the calculation of the mean force, the only 
part which is of interest is 
the {\it renormalized} expectation 
value, which is obtained when ${\tilde G}_{\mu\nu'}$
rather than $G_{\mu\nu'}$ is used in Eq.~(\ref{eq:E2}). This is 
simply subtracting 
out the pure vacuum contribution, and is same as normal ordering with respect to 
the Minkowski 
vacuum. Equation (\ref{eq:E2}) can be rewritten as 
\begin{equation}
  \langle E_i\,E_{j'}\rangle = \langle :E_i\,E_{j'}:\rangle 
                               +\langle E_i\,E_{j'}\rangle_{0} \, , 
\end{equation}
where the normal-ordered term is
\begin{equation}
\langle :E_i\,E_{j'}:\rangle =
\partial_0 \,\partial_{0'} \,{\tilde G}_{ij'}+
\partial_i \,\partial_{j'} \,{\tilde G}_{00'}-
\partial_0 \,\partial_{j'} \,{\tilde G}_{i0'}-
\partial_i \,\partial_{0'} \,{\tilde G}_{0j'}
\label{eq: E2_NO}
\end{equation}
and the vacuum term is
\begin{equation}
\langle E_i\,E_{j'}\rangle_{0} =
\partial_0 \,\partial_{0'} \,G^{(0)}_{ij'}+
\partial_i \,\partial_{j'} \,G^{(0)}_{00'}-
\partial_0 \,\partial_{j'} \,G^{(0)}_{i0'}-
\partial_i \,\partial_{0'} \,G^{(0)}_{0j'} \,.
\label{eq: E2_cross}
\end{equation}
If we combine
Eqs.~(\ref{eq:force}), (\ref{eq:Gtilde}), and (\ref{eq: E2_NO}), we 
obtain
the Casimir-Polder result for the mean force:
\begin{equation}
\langle {\bf :F:} \rangle = -{3\alpha\over{8\pi^2}}\,
{\hbar c\over z^5}\;\hat z.
\label{eq:MeanForce}
\end{equation}
This is an attractive force in the direction  perpendicular to the 
conducting plate.

Recall that Eq.~(\ref{eq:MeanForce}) strictly holds only when the particle
is described by a static (frequency-independent) polarizability. For the
case of a one electron atom in its ground state, Casimir and Polder gave
a more complicated expression which reduces to Eq.~(\ref{eq:MeanForce})
in the large $z$ limit. In the case of a macroscopic particle with nontrivial
dispersive properties, there is the possibility of having a force which is
either attractive or repulsive, and larger in magnitude than that given
by  the above expression \cite{Ford98}. In the present paper, we will
deal only with the case of a frequency-independent polarizability.

\section{The Force-Force Correlation Function} \label{sec: force 
correlation}

    Now we wish to study the
fluctuations in this force. This may be done by examining the 
correlation function 
\(\langle :\bf{F(x)}::\bf{F(x')}: \rangle\) and the 
expectation
value of the squared force $\langle :\bf{F}:^{2}\rangle$. However, we 
will encounter 
the quantity $\langle{:\bf E}^2(x)::{\bf E}^2(x'):\rangle$, which is 
formally 
divergent in the coincident limit $x'\to x$.
Unlike the quadratic expectation values encountered in the case of 
the mean force,
we cannot simply render this quantity finite by subtracting its 
expectation value in
the Minkowski vacuum state. Following the method used in our previous 
works 
\cite{WF99,WF00}, this two point function can be
decomposed into three different terms by using Wick's theorem
\begin{equation}
\langle :{\bf E}^2(x)::{\bf E}^2(x'):\rangle
  =\langle :{\bf E}^2(x){\bf E}^2(x'): \rangle
  +\langle :{\bf E}^2(x)::{\bf E}^2(x'):\rangle_{cross}
  +\langle :{\bf E}^2(x)::{\bf E}^2(x'):\rangle_{0} \, ,
    \label{eq:}
\end{equation}
which are the fully normal-ordered term, the cross term and 
the pure vacuum term, respectively. In the coincidence limit \(x\to 
x'\), the fully normal-ordered 
term is a well-defined local quantity. The cross term contains a 
state-dependent 
divergence, but can be made finite with careful regularization in the 
integral.
The pure vacuum term is also divergent in the coincidence limit, but
it is  state-independent and cancels when we measure the difference due 
to a changes of the boundary condition.
The fully normal-ordered term can be expressed explicitly as
\begin{eqnarray}
\langle :{\bf E}^2(x){\bf E}^2(x'): \rangle & = &
\langle :{E_i E^i\,E_{j'}E^{j'}}: \rangle \nonumber \\
 & = & \langle :E_i\,E^i: \rangle\, 
\langle :E_{j'}\,E^{j'}:\rangle + \langle :E_i\,E_{j'}:\rangle\,
\langle :E^i\,E^{j'}:\rangle + \langle :E_i\,E^{j'}:\rangle\,
\langle :E^i\,E_{j'}:\rangle \,
\label{eq: n.o.}
\end{eqnarray}
and the cross term is 
\begin{equation}
 \langle :{\bf E}^2(x)::{\bf E}^2(x'):\rangle_{cross}
 =\langle :E_i E^i:\,:E_{j'}E^{j'}:\rangle_{cross}
 =4\langle :E_{i}E_{j'}:\rangle \langle E^{i}E^{j'}\rangle_{0}\, .
    \label{eq: cross}
\end{equation}
The physical content of both of these terms has been discussed by us 
\cite{WF99,WF00} in other contexts. In general, both terms can
contribute to the fluctuations of the stress tensor, or other quadratic 
operators. 

The force-force correlation function $\langle :F_i:: F_{j'}: 
\rangle$, 
evaluated at ${\bf x} = {\bf x'}$ but at
different times, can be obtained by the formula
\begin{equation}
\langle :F_{k}(t,z)::\,F_{k}(t',z): \rangle 
= \frac{\alpha^{2}}{4}\biggl[ \partial_k\,\partial_{k'}\,
\langle{:E_i E^i::\,E_{j'}E^{j'}:}\rangle \biggr]_{{\bf x}= {\bf x'}} 
\,.
\end{equation}
Again, this correlation function contains two parts we are interested 
in, namely
the fully normal-ordered term and the cross term, and the cross term 
is divergent 
in the coincident limit, \( t \to t' \). The contribution from these 
two terms 
will be examined separately in the following section. The idea 
is to 
investigate the velocity dispersion of a test particle due to this 
fluctuating
Van der Waals force.

\section{Velocity Fluctuations of  a test particle}\label{sec:average}

We can better understand the effects of these fluctuations by
studying the motion of particles subjected to the fluctuating force, 
which will be described by a Langevin equation. 
Consider particles which start at rest 
at time $t=0$. The mean velocity at a later time \(t\) is given by
\begin{equation}
\langle :v_k(t):\rangle=
{1\over m}\,\int_0^t dt_1\,\langle :F_k(t_1,z):\rangle \, ,
\label{eq:V}
\end{equation}
where \(k=x,y,z\). To simplfy the analysis, we assume that the distance 
of the particle from the plate does not change significantly in a time $t$, 
so that $z$ is approximately constant.
Then the dispersion around the mean velocity in the $k$-direction at 
a later
time is given by
\begin{equation}
\langle \triangle v_k^2(t)\rangle=
{1\over m^2}\,\int_0^t dt_1\,\int_0^t dt_2\,
\Bigl[\langle :F_k(t_1,z):\,:F_k(t_2,z):\rangle 
-\langle :F_k(t_1,z):\rangle\, \langle :F_k(t_2,z):\rangle \Bigr] \, .
\label{eq:V^2}
\end{equation}
This can be decomposed into two terms 
\begin{equation}
    \langle\triangle v_{k}^{2}\rangle
    =\langle :\triangle v_{k}^{2}:\rangle 
    + \langle \triangle v_{k}^{2}\rangle_{cross}\, ,
    \label{eq: v^2_cross}
\end{equation}
which are the fully normal-ordered term
\begin{equation}
    \langle :\triangle v_{k}^{2}(t):\rangle
    ={1\over m^2}\,\int_0^t dt_1\,\int_0^t dt_2\,
\Bigl[\langle :F_k(t_1,z)\,F_k(t_2,z):\rangle 
-\langle :F_k(t_1,z):\rangle\, \langle :F_k(t_2,z):\rangle \Bigr]
\label{eq: v^2_NO}
\end{equation}
and the cross term
\begin{equation}
    \langle \triangle v_{k}^{2}(t)\rangle_{cross}
    ={1\over m^2}\,\int_0^t dt_1\,\int_0^t dt_2\,
\langle :F_k(t_1,z):\,:F_k(t_2,z):\rangle_{cross} \, .
\label{eq: v2-1}
\end{equation}
Here  the pure vacuum term is dropped 
because we are only 
interested in the difference due to a change of boundary 
conditions, which is the 
change caused by adding a plate. The fully normal-ordered term 
and the cross term will now be discussed in turn.

\subsubsection{The fully normal-ordered term}\label{subsubsec:n.o.}

   Consider the force fluctuations due to the fully normal-ordered term, 
Eq.~(\ref{eq: n.o.}). 
The diagonal components of the force-force correlation functions are 
\begin{equation}
\langle :F_{k}(t,z)\,F_{k}(t',z): \rangle 
= \frac{\alpha^{2}}{4}\biggl[ \partial_k\,\partial_{k'}\,
\langle{:E_i E^i\,E_{j'}E^{j'}:}\rangle \biggr]_{{\bf x}= {\bf x'}} 
\,.
                                           \label{eq:f2def}
\end{equation}
Note that the off-diagonal terms will be zero in the limit, 
\(\bf{x} \to \bf{x'}\).
Use Eq. (\ref{eq:Gtilde}) and Eq. (\ref{eq: E2_NO}), and we find that 
the electric 
field two point functions can be expressed as
\begin{eqnarray}
    \langle :E_{x}E_{x'}:\rangle & = & 
    \frac{1}{2\pi^{2}}\biggr[\frac{2}{\tilde D^{2}}
    -\frac{4(x-x')^{2}}{\tilde D^{3}}
    +\frac{4(t-t')^{2}}{\tilde D^{3}}\biggr]
    \label{eq:EE_x}  \\
    \langle :E_{y}E_{y'}:\rangle  & = & 
    \frac{1}{2\pi^{2}}\biggr[\frac{2}{\tilde D^{2}}
    -\frac{4(y-y')^{2}}{\tilde D^{3}}
    +\frac{4(t-t')^{2}}{\tilde D^{3}}\biggr]
    \label{eq:EE_y}
\end{eqnarray}
and
\begin{equation}
     \langle :E_{z}E_{z'}:\rangle 
     = \frac{-1}{2\pi^{2}}\biggr[\frac{2}{\tilde D^{2}}
    -\frac{4(z+z')^{2}}{\tilde D^{3}}
    +\frac{4(t-t')^{2}}{\tilde D^{3}}\biggr] \,.
    \label{eq:EE_z}
\end{equation}
Plug Eqs.~(\ref{eq:EE_x}), (\ref{eq:EE_y}) and 
(\ref{eq:EE_z}) into 
 Eqs.~(\ref{eq: n.o.}) and (\ref{eq:f2def}). We find
\begin{equation}
 \langle :F_x(t,z)\,F_x(t',z): \rangle  
 =\langle :F_y(t,z)\,F_y(t',z): \rangle 
 =-{{4\alpha^2\left(5\,{T^4} + 16\,{T^2}\,{z^2} + 48\,{z^4}\right)}
\over{\pi^4\left( T^2-4z^2\right)^7 }}   
    \label{eq:f2_x}
\end{equation}
and
\begin{equation}
\langle :F_z(t,z)\,F_z(t',z): \rangle 
-\langle :F_{z}(t,z):\rangle \langle :F_{z}(t',z):\rangle = 
{{4\alpha^2\left(5\,{T^6} + 
308\,{T^4}\,{z^2} + 944\,{T^2}\,{z^4} + 
1728\,{z^6}\right)}
\over{\pi^4\left( T^2-4z^2\right)^8 }} \, ,
                                           \label{eq:f2_z}
\end{equation}
where $T = t-t'$ and the product of the mean force is 
\begin{equation}
\langle :F_z(t,z):\rangle \langle :F_z(t',z): \rangle 
= \frac{\alpha^{2}}{4} \biggl[
\partial_z\langle{:E_i E^i:\rangle \,\partial_{z'}
\langle :E_{j'}E^{j'}:}\rangle \biggr]_{{\bf x}= {\bf x'}} \,.
                                           \label{eq:nf2def}
\end{equation}
All of these results are independent of x and y, and are  
Lorentz invariant under boosts in the directions parallel to the plate.
In the limit \(t'\to t \), these fluctuations become
\begin{equation}
    \langle :\Delta F^{2}_{x}(z,t):\rangle
    = \langle :\Delta F^{2}_{y}(z,t):\rangle 
    = \frac{3\,\alpha^{2}}{256\, \pi^{4}\, z^{10}}
    \label{eq:f2^x}
\end{equation}
and
\begin{equation}
    \langle :\Delta F^{2}_{z}(z,t):\rangle  =  
    \frac{27\,\alpha^{2}}{256\, \pi^{4}\, z^{10}} \, .
    \label{eq:f2^z}
\end{equation}
Here the z component is about 5 times  the x and y components.
If we compare the expectation value of the squared force
from Eq.~(\ref{eq:f2^z}) with the square of the expectation value from
Eq.~(\ref{eq:MeanForce}), we obtain a measure of the force 
fluctuations:
\begin{equation}
\Delta=\Biggl|{{\langle :F_z(t,z)^2:\rangle-
\langle :F_z(t,z):\rangle ^2}
\over{\langle :F_z(t,z):\rangle ^2}}\Biggr|
={3\over 4}\; ,
\end{equation}
which is of order of unity and shows that the force is fluctuating 
considerably. Note that even though there are no mean forces in x and y 
directions, the
deviation of the force in these directions are still non-zero.  
Furthermore, the correlation function,  Eq.~(\ref{eq:f2_x}) and 
 Eq.~(\ref{eq:f2_z}),
becomes small if $T\gg z$, i.e. for time separations large compared to
the distance of the atom from the plate. This shows that the 
characteristic
fluctuation time is of the order of $z$. 

However, the behavior at time scales larger than the characteristic 
fluctuation time \(z\) is also important, and is needed to find the
velocity fluctuations. Note that
the force correlation functions, Eqs.~(\ref{eq:f2_x}) and ~(\ref{eq:f2_z}), 
are singular at $T=2z$, a time separation equal to the round-trip time light 
travel between the particle and the plate. This singularity is presumably an
artifact of our assumption of a perfectly reflecting plate, and would hence
be smeared out in a more realistic treatment. However, we will see that the 
integrals can be made well-defined even with this singular integrand.

Let us now change integration variables to $T=t_1 -t_2$ and $\tau= 
t_1 +t_2$.
If $F(T)$ is an even function, then
\begin{equation}
\int_0^t dt_1\,\int_0^t 
dt_2\,F(T) = 
\frac{1}{2}\int_{-t}^t dT\,\int_{|T|}^{2t-T} d\tau \,F(T) =
2\int_{0}^t dT\,(t-T)\,F(T) \,. 
\end{equation}
With Eqs.~(\ref{eq:MeanForce}),  (\ref{eq:f2_x}) and (\ref{eq:f2_z}),
 we may write Eq.~(\ref{eq: v^2_NO}) as
\begin{equation}
 \langle :\Delta v_k^2(t):\rangle =
{1\over m^2}\,\int_0^t dT \frac{f_{k}(T)}{(T-2z)^8}\, ,  
\label{eq:Delv^2'}
\end{equation}
where 
\begin{equation}
    f_{x}(T)
    =f_{y}(T)
    =-{{8\,{{\alpha }^2}\,\left( 5\,{T^4} + 
     16\,{T^2}\,{z^2} + 48\,{z^4} \right) (t-T)}\over 
{{{\pi}^4}\,{{\left({T}+2\,z\right)}^7}}}
\end{equation}
and
\begin{equation}
f_{z}(T)=
{{8\,{{\alpha }^2}\,\left( 5\,{T^6} + 
308\,{T^4}\,{z^2} + 944\,{T^2}\,{z^4} + 
1728\,{z^6} \right) (t-T)}\over 
{{{\pi}^4}\,{{\left({T}+2\,z\right)}^8}}}\, .
\end{equation}
 The dispersion 
$ \langle :\Delta v_k^2(t):\rangle$ in Eq.~(\ref{eq:Delv^2'}) will be 
defined  as a generalized principle value integral \cite{Davies}. 
Such integrals involving higher-order 
poles may be evaluated by successive integrations by parts, which remove 
the divergence 
at the point \(T=2z\) and lead to the formula
\begin{eqnarray}
\wp\int_a^b dx\,\,{f(x)\over (x-c)^n}
&=& -{1\over{(n-1)!}}\,\left[\sum_{i=0}^{n-2}
(n-2-i)!\,f^{(i)}(x)\,(x-c)^{-n+1+i}\right]_a^b
\nonumber\\
&&+{1\over{(n-1)!}}\,\wp\int_a^b dx\,
f^{(n-1)}(x)\,(x-c)^{-1} . \nonumber\\
\end{eqnarray}
We may now apply this formula to evaluate $ \langle :\Delta 
v_k^2(t):\rangle$.
The result simplifies considerably if we assume that $t \gg 2z$, 
which is
the limit of greatest physical interest. In this case, after some 
calculation
one finds that
\begin{equation}
  \langle :\Delta v_x^2(t,z):\rangle
  =\langle :\Delta v_y^2(t,z):\rangle
  = {1\over{m^2}}\,\left({9\over {1280 \pi^4}} \right)\,
    {{\alpha^2 \hbar^2}\over{z^8}}\, + O\left( \frac{z}{t}\right) 
    \label{eq:MainResult_x,y}
\end{equation}
and
\begin{equation}
 \langle :\Delta v_z^2(t,z):\rangle
= {1\over{m^2}}\,\left({63\over {1280 \pi^4}} \right)\,
    {{\alpha^2 \hbar^2}\over{z^8}}\,
   + O\left( \frac{z}{t}\right) \, .
   \label{eq:MainResult_z}
\end{equation}
Note that even though there is no mean force in the x and y directions, the 
dispersion of the velocity in the direction parallel to the plate is still 
nonzero.

\subsubsection{The cross term}\label{subsubsec:cross}

The other intriguing part of the quantum fluctuation of the Van der 
Waals force is the cross term. Its contribution to the velocity fluctuation,
Eq.~(\ref{eq: v2-1}), is formally divergent. However, it can be made finite
by an integration by parts procedure analogous to that used in the previous 
section. The key assumption which we need to introduce is one of
{\it adiabatic switching}. This means that the effect of the plates is smoothly 
switched on in the past and then off in the future. Physically, this might be 
achieved by means of a plate whose reflectivity could be controlled.
This switching will allow us to drop surface terms which would otherwise
be divergent. An analogous switching was assumed in a treatment of the
quantum fluctuations of radiation pressure \cite{WF00}. There it was shown 
that the cross term plays a central role, and in fact gives the sole
contribution when a laser beam in a coherent state is shined on a mirror.
In this case, it was necessary to assume that the laser beam is 
switched on in the past and then off in the future in order to obtain
finite velocity fluctuations for the mirror.

In analogy to Eq.~(\ref{eq:f2def}), the cross term of these 
force-force two point functions is defined by
\begin{equation}
\langle F_k(t,z)\,F_k(t',z) \rangle_{cross} 
= \frac{\alpha^{2}}{4}\biggl[ \partial_k\,\partial_{k'}\,
\langle{:E_i E^i:\,:E_{j'}E^{j'}:}\rangle_{cross} \biggr]_{{\bf x}= 
{\bf x'}} \,.
 \label{eq:f2_cross}
\end{equation}
Due to Eq. (\ref{eq: cross}), we need to know the vacuum two point 
function 
$\langle E_{i}E_{j'}\rangle_{0}$ as well as the normal-ordered two 
point function
$\langle :E_{i}E_{j'}:\rangle$ to compute the cross term. Use the 
equations 
(\ref{eq:G_0}) and (\ref{eq:D_0}) to compute these vacuum
two-point functions Eq. (\ref{eq: E2_cross}), and find
\begin{eqnarray}
    \langle E_{x}E_{x'}\rangle_{0} & = & 
    \frac{-1}{2\pi^{2}}\biggr[\frac{2}{ D^{2}}
    -\frac{4(x-x')^{2}}{ D^{3}}
    +\frac{4(t-t')^{2}}{ D^{3}}\biggr]
    \label{eq:EE_x0}  \\
    \langle E_{y}E_{y'}\rangle_{0}  & = & 
    \frac{-1}{2\pi^{2}}\biggr[\frac{2}{ D^{2}}
    -\frac{4(y-y')^{2}}{ D^{3}}
    +\frac{4(t-t')^{2}}{ D^{3}}\biggr]
    \label{eq:EE_y0}
\end{eqnarray}
and
\begin{equation}
     \langle E_{z}E_{z'}\rangle_{0} 
     = \frac{-1}{2\pi^{2}}\biggr[\frac{2}{ D^{2}}
    -\frac{4(z-z')^{2}}{ D^{3}}
    +\frac{4(t-t')^{2}}{ D^{3}}\biggr] \,.
    \label{eq:EE_z0}
\end{equation}
Plug these vacuum two point functions, along with the normal-ordered 
two point 
functions Eqs.~(\ref{eq:EE_x}), (\ref{eq:EE_y}) and 
 (\ref{eq:EE_z}) into Eq. (\ref{eq: cross}) and compute the derivatives 
in Eq. (\ref{eq:f2_cross}). The force-force two point functions then becomes
\begin{equation}
\langle F_k(t,z)\,F_k(t',z) \rangle_{cross}  
= \frac{\alpha^{2}}{\pi^{2}} 
  \biggr[\frac{f_{k,4}}{(t_{1}-t_{2})^{4}} - \frac{4\, 
f_{k,6}}{(t_{1}-t_{2})^{6}}\biggr]\,,
\label{eq:f-f_k}
\end{equation}
Here $f_{k,4}$ and $f_{k,6}$ are formed from the normal-ordered 
two point 
functions and their derivatives, and can be expressed as 
\begin{equation}
f_{k,4}=
\biggl[ \partial_{k,k'}\langle :E_{x}(x)E_{x'}(x'):
\rangle + \partial_{k,k'}\langle :E_{y}(x)E_{y'}(x'):
\rangle +  \partial_{k,k'}\langle :E_{z}(x)E_{z'}(x'):
\rangle \biggr]_{{\bf x}= {\bf x'}}
    \label{eq:f4}
\end{equation}
and as
\begin{equation}
f_{x,6}=
\biggl[ \langle :E_{x}(x)E_{x'}(x'):
\rangle + 2\langle :E_{y}(x)E_{y'}(x'):
\rangle +  2\langle :E_{z}(x)E_{z'}(x'):
\rangle \biggr]_{{\bf x}= {\bf x'}} \, ,
    \label{eq:f6-x}
\end{equation}
\begin{equation}
f_{y,6}=
\biggl[ 2\langle :E_{x}(x)E_{x'}(x'):
\rangle + \langle :E_{y}(x)E_{y'}(x'):
\rangle +  2\langle :E_{z}(x)E_{z'}(x'):
\rangle \biggr]_{{\bf x}= {\bf x'}} 
    \label{eq:f6-y}
\end{equation}
and
\begin{equation}
f_{z,6}=
\biggl[ 2\langle :E_{x}(x)E_{x'}(x'):
\rangle + 2\langle :E_{y}(x)E_{y'}(x'):
\rangle +  \langle :E_{z}(x)E_{z'}(x'):
\rangle \biggr]_{{\bf x}= {\bf x'}} \, .
    \label{eq:f6-z}
\end{equation}
The singular parts, \(1/(t_{1}-t_{2})^{4}\) and 
\(1/(t_{1}-t_{2})^{6}\), in 
Eq. (\ref{eq:f-f_k}) are caused
by the vacuum two point functions.
Use Eq. (\ref{eq:f-f_k}) and change the variables ($t_{1}$,$t_{2}$) to 
dimensionless 
ones 
($s_{1}=t_{1}/2z$,$s_{2}=t_{2}/2z$). 
The velocity fluctuation Eq.~(\ref{eq: v2-1}) becomes
\begin{equation}
    \langle \triangle v_{k}^{2}(t)\rangle_{cross}
    =\frac{\alpha^{2}}{\pi^{2}m^{2}}\int_{0}^{t/2z}\int_{0}^{t/2z}
    \biggr[\frac{f_{k,4}}{(2z)^{2}(s_{1}-s_{2})^{4}}
    -\frac{4\, f_{k,6}}{(2z)^{4}(s_{1}-s_{2})^{6}}\biggr]
    \, ds_{1}ds_{2} \, .
    \label{eq: v_cross_f}
\end{equation}
Because of the adiabatic switching assumption discussed above, we can now 
integrate by parts and drop the surface terms, using the relations
\begin{equation}
    \int\int \frac{f_{k,4}}{(s_{1}-s_{2})^{4}}ds_{1}ds_{2}
    =-\frac{1}{12}\int\int \biggl[ 
(\partial_{s_{1}})^{2}(\partial_{s_{2}})^{2} 
    f_{k,4}\biggr]\, \ln (s_{1}-s_{2})^{2} \, ds_{1}ds_{2}
\end{equation}
and
\begin{equation}
    \int\int \frac{f_{k,6}}{(s_{1}-s_{2})^{6}}ds_{1}ds_{2}
    =\frac{1}{240}\int\int \biggl[ 
(\partial_{s_{1}})^{3}(\partial_{s_{3}})^{3} 
    f_{k,6}\biggr]\, \ln (s_{1}-s_{2})^{2} \, ds_{1}ds_{2} \, .
\end{equation}
Plug these re-defined integrals into Eq. (\ref{eq: v_cross_f}) and 
change variables to
($u=s_{1}-s_{2}$, $v=s_{1}+s_{2}$). Use of the relation
\begin{equation}
    \int_{0}^{a}\int_{0}^{a}\, ds_{1}ds_{2}
    =\frac{1}{2}\biggl( \int_{-a}^{0}du\int_{-u}^{u+2a}dv
    +\int_{0}^{a}du\int_{u}^{2a-u}dv\biggr) \, ,
\end{equation}
leads to 
\begin{eqnarray}
     \langle \triangle v_{k}^{2}(t)\rangle_{cross} 
     & = & 
\frac{\alpha^{2}}{\pi^{2}m^{2}}\int_{0}^{t/2z}\int_{0}^{t/2z}
    g_{k}(s_{1},s_{2})\, \ln (s_{1}-s_{2})^{2} \, ds_{1}ds_{2}
    \nonumber  \\
     & = & \frac{\alpha^{2}}{\pi^{2}m^{2}}
     \biggl( 2t\int_{0}^{t/2z}\, g_{k}(u^{2})\, \ln u^{2}\, du 
     -2\int_{0}^{t/2z} \, u\, g_{k}(u^{2})\, ln\, u^{2} \, du 
     \biggr)\, ,
    \label{eq:v_cross_uv}
\end{eqnarray}
where
\begin{equation}
    g_{x}=g_{y}=-\frac{3(5+275u^{2}+1325u^{4}+1041u^{6}+42u^{8})}
    {16\pi^{2}(u^{2}-1)^{9}z^{8}}
\end{equation}
and
\begin{equation}
    g_{z}=\frac{3(23+663u^{2}+1573u^{4}+429u^{6})}
    {8\pi^{2}(u^{2}-1)^{9}z^{8}}\, .
\end{equation}
When $t>>2z$, the first term in Eq. (\ref{eq:v_cross_uv}) goes to 
zero for all the 
components \(k=x,y,z\), which leads to
\begin{equation}
    \langle \triangle v_{x}^{2}(z)\rangle_{cross}
    =\langle \triangle v_{y}^{2}(z)\rangle_{cross}
    \to \frac{13}{240}\frac{\hbar^2 \,\alpha^{2}}{\pi^{4}m^{2}z^{8}}
    \label{eq:v_cross_final_x,y}
\end{equation}
and
\begin{equation}
    \langle \triangle v_{z}^{2}(z)\rangle_{cross}
    \to -\frac{497}{480}\frac{\hbar^2 \,\alpha^{2}}{\pi^{4}m^{2}z^{8}}\, .
    \label{eq:v_cross_final_z}
\end{equation}
The magnitude of the velocity fluctuation of the z component is about 
19 times of that
of x and y components. In all cases, 
the contributions of the cross terms are larger than those of the 
fully normal-ordered terms, 
Eqs.~(\ref{eq:MainResult_x,y}) and  (\ref{eq:MainResult_z})
\begin{eqnarray}
    \langle\triangle v_{x}^{2}\rangle_{cross}=\langle\triangle 
v_{y}^{2}\rangle_{cross} 
    &  \approx
    &   7.7\, \langle :\triangle v_{x}^{2}:\rangle
    \nonumber \\
    \langle\triangle v_{z}^{2}\rangle_{cross} & \approx 
    & -21\, \langle :\triangle v_{z}^{2}:\rangle \, .
    \label{eq:ratio}
\end{eqnarray}
The most surprising result is the negative z-component due to 
the cross term.
The total velocity fluctuations are
\begin{equation}
   \langle\triangle v_{y}^{2}\rangle
   =\langle\triangle v_{x}^{2}\rangle
   =\langle :\triangle v_{x}^{2}:\rangle + \langle\triangle 
v_{x}^{2}\rangle_{cross}
   =\frac{47}{768}\frac{\hbar^2 \,\alpha^{2}}{\pi^{4}m^{2}z^{8}}
    \label{eq:final_x} 
\end{equation}
and
\begin{equation}
   \langle\triangle v_{z}^{2}\rangle
   =\langle :\triangle v_{z}^{2}:\rangle + \langle\triangle 
v_{z}^{2}\rangle_{cross}
   =-\frac{3787}{3840}\frac{\hbar^2 \,\alpha^{2}}{\pi^{4}m^{2}z^{8}}
    \label{eq:final_z} \, .
\end{equation}
The results are independent of time and the z 
component is still negative after the fully normal-ordered term is 
added to the cross term.
The time-independent result shows a behavior similar to the case 
of Brownian 
motion in thermal equilibrium system. However we should also note 
that the velocity dispersion is not isotropic. The x and y components are much 
smaller than the magnitude of the $z$ component 
\begin{equation}
    \langle\triangle v_{x}^{2}\rangle
    =\langle\triangle v_{y}^{2}\rangle
    \approx 0.06 |\langle\triangle v_{z}^{2}\rangle |\, .
\end{equation}
The non-isotropic behavior is also reflected in the 
fact that the mean force is zero in the parallel direction, but 
non-zero in the perpendicular direction. 
That the $\langle\triangle v_{i}^{2}\rangle$ approach constant values,
as opposed to growing in time, can be understood on the basis of energy
conservation. 

Of particular interest is the fact that $\langle\triangle v_{z}^{2}\rangle < 0$.
Recall that this quantity is a difference between a mean squared velocity
with the plate and one without it, hence it is possible for this difference 
to be negative. (Similarly, the negative energy density in the Casimir effect
arises from energy density being defined as a difference.) However, this
negative value requires a physical interpretation. The most plausible
explanation is that one cannot ignore the quantum nature of the test particles
we have been discussing. The particle must have both a position uncertainty
\(\Delta z\) and a momentum uncertainty \(\Delta p_z\), obeying the uncertainty 
principle. Furthermore, because the particle is massive, there will be
wavepacket spreading in which \(\Delta z\) is an increasing function of time.
Thus, even if the particle is initially in a minimum uncertainty
wavepacket, at later time it will satisfy the uncertainty principle by
a wide margin. Our interpretation of the negative 
$\langle\triangle v_{z}^{2}\rangle$ is that the electromagnetic vacuum
fluctuations cause a small reduction in the velocity spread of the wavepacket
compare to what it would have been without the plate present. Imagine that
we initially prepare the particle in a minimum uncertainty state, and then
allow it to evolve for a time \(\tau >> 2z\)).  During
this time \(\Delta z\) will increase, but \(\Delta v_z\) will decrease slightly.

In any case, the magnitude of the velocity changes due to
electromagnetic vacuum fluctuations is always very small compared to the
velocity spread due to quantum uncertainty. Let the latter be
\begin{equation}
\Delta v_q = \frac{\Delta p_z}{m} > \frac{\Delta z}{m} \, .
\end{equation}
Compare this to the spread due to vacuum fluctuations,
\begin{equation}
\Delta v_f = {\rm max}(\sqrt{|\langle\triangle v_{i}^{2}\rangle|})
           \approx \frac{\alpha}{\pi^2 m z^4}\, .
\end{equation}
Their ratio satisfies
\begin{equation}
\frac{\Delta v_f}{\Delta v_q} < \left(\frac{\Delta z}{z}\right)
                               \left(\frac{\alpha}{\pi^2 \, z^3}\right)\, .
                                         \label{eq:vratio}
\end{equation}
However, both factors on the right-hand-side of the above expression
are small compared to one. The particle must be localized in a region small
compared to the distance to the plate, so $\Delta z \ll z$. The size of
the particle must also be small compared to $z$, and because the 
polarizability $\alpha$ is at most of the order of the volume of the particle,
$\alpha \ll z^3$. Thus
\begin{equation}
\frac{\Delta v_f}{\Delta v_q} \ll 1 \, .
\end{equation}

\section{Discussion }\label{sec:discuss}

      Equations~(\ref{eq:final_x}) and ~(\ref{eq:final_z}) 
tell us that the effect of the
fluctuations of the retarded Van der Waals force is to generate a
random motion around that described by the classical trajectory. Of
course, if a particle is released in the vicinity of a conducting 
plate,
it tends to fall toward the plate under the influence of the mean 
force,
Eq.~(\ref{eq:force}). However, we could apply a compensating classical
force ${\bf F}_{cl} = -\langle {\bf F} \rangle$, 
so that the classical trajectory is that of a particle at fixed 
$z$. Nonetheless, it will still develop a mean squared velocity given 
by Eqs.~(\ref{eq:MainResult_x,y}) and (\ref{eq:MainResult_z}). 
If we look at the $x$-direction (or \(y\)-direction), this is 
equivalent to thermal motion at a effective temperature of
\begin{equation}
T_{eff} = \left({47\over {768 \pi^4}} \right)\, {{\alpha^2 
\hbar^2}\over
{k_B\,m\,z^8}}\, ,   \label{eq:Teff}
\end{equation}
where $k_B$ is Boltzmann's constant.
Equation~(\ref{eq:Teff}) can be written as
\begin{equation}
T_{eff} \approx 10^{-1} K \left(\frac{m_H}{m}\right)
\left(\frac{1 \AA}{z}\right)^8 
\left(\frac{\alpha}{\alpha_H}\right)^2\,,   
\end{equation}
where $m_H$ and $\alpha_H$ are the mass and static polarizability of atomic
hydrogen, respectively. This effective temperature is essentially
the temperature below which the system must be cooled so that the
quantum fluctuation effects are not masked by ordinary thermal fluctuations.
The effect in the $z$-direction is different from that in the transverse
directions in that the mean squared velocity in that direction is
reduced. Nonetheless, Eq.~(\ref{eq:Teff}) gives an estimate of the magnitude
even in this case.
The magnitude of the effect depends crucially upon how small $z$ can
be. For atoms near a metal plate, both the assumptions of perfect
conductivity and of using the static (as opposed to dynamic 
polarizability)
break down for sufficiently small $z$, typically for $z \lesssim 10^3 
\AA$. 
Thus the effect of the fluctuations will be very small in the range
that both of these asumptions hold well. However, there is likely to be
some effect even at much smaller values of $z$. A metal surface acts
as a partial reflector of electromagnetic waves even up into the x-ray
range, where Bragg scattering can produce reflectivities close to $100 \%$
at special angles \cite{CS01}.
  Thus although Eq.~(\ref{eq:Teff}) is strictly valid only for
$z > 10^3 \AA$, it may produce crudely correct answers for $z$ as small
as as a few \AA. If so, the fluctuation effects could conceivably approach   
observable levels. This conjecture needs to be confirmed by more detailed
treatements.

The appearance of nonzero values for $\langle\triangle v_{x}^{2}\rangle$
and $\langle\triangle v_{y}^{2}\rangle$ requires some comment. By symmetry,
a particle is equally likely to be deflected by the electromagnetic 
field in the $+x$ or $-x$ directions, and hence $\langle v_x \rangle
= \langle v_y \rangle = 0$. However, the history of an individual particle
does not have to respect the symmetry of the problem. Some particles 
acquire nonzero transverse components of velocity, leading to 
$\langle\triangle v_{x}^{2}\rangle \not= 0$. A similar situation arises
in lightcone fluctuations due to quantum gravity effects in a compact 
space \cite{YF}. Here Lorentz invariance holds on the average, but not 
for the history of an individual test particle.

The time scale of the fluctuations due to the fully normal-ordered term are 
of the order of $z$, the light travel time between the particle and the plate,
as may be seen from the fact that the correlation functions, 
Eqs.~(\ref{eq:f2_x}) and (\ref{eq:f2_z}), vanish for $T \gg z$. The time
scale associated with the fluctuations arising from the cross term is
of the same order. The short distance singularity of the cross term
indicates that it contains fluctuations on arbitrarily short scales. However,
these very rapid fluctuations are averaged out by the time integrations.
The final integral for $\langle \triangle v_{k}^{2}(t)\rangle_{cross}$,
Eq.~(\ref{eq:v_cross_uv}), again contains an integrand which vanishes rapidly
for $u = T/(2z) \ll 1$.

In summary, the Casimir-Polder result is a mean force, whereas the actual
 force is rapidly fluctuating. The typical magnitude of the
fluctuations is of the same order as the mean force itself, but the 
time
scale of the fluctuations is of the order of the light travel time 
between
the atom and the plate. For most purposes, such as the Sukenik et al 
\cite{Sukenik} experiment, the fluctuations average to zero and are 
not
seen. In principle, it is possible to detect the fluctuations through
the random motions which they will induce in test particles. 
For ordinary atomic systems, this effect is very small.  

The effect discussed in this paper is also of interest in gravity
theory. When quantum matter fields act as the souce of gravity,
fluctuations of the stress tensor will lead to ``passive'' fluctuations
of the spacetime geometry. These fluctuations are one of the physical
phenomena to be expected in any quantum theory of gravity. 
 Quantum fluctuations of the spacetime metric imply Brownian
motion of the test particles which probe the fluctuating metric \cite{Kuo,WF01}.
Thus the Brownian motion due to electromagnetic vacuum fluctuations
treated here is a useful analogy for understanding the quantum nature
of gravity.

\vspace{0.5cm}
{\bf Acknowledgement}: This work was supported by the National Science
Foundation under Grant PHY-9800965.

\end{document}